\documentclass[%
reprint,
nofootinbib,
superscriptaddress,
 amsmath,
 amssymb,
 aps,
]{revtex4-1}

\usepackage{cancel}
\usepackage{graphicx}
\usepackage{dcolumn}
\usepackage{bm}
\usepackage{hyperref}
\usepackage[mathlines]{lineno}
\usepackage{color}
\usepackage{braket}
\usepackage[normalem]{ulem}
\usepackage[english]{babel}
\definecolor{dgreen}{rgb}{0,0.39,0}

\begin{document}

\preprint{APS/123-QED}

\title{Relativistic Generalized Uncertainty Principle}

\author{Vasil Todorinov}
 \email{v.todorinov@uleth.ca}
\affiliation{Theoretical Physics Group and Quantum Alberta,
Department of Physics and Astronomy,
University of Lethbridge,
4401 University Drive, Lethbridge,
Alberta, T1K 3M4, Canada}
\author{Pasquale Bosso}
\email{bosso@fisica.ugto.mx}
 \affiliation{Divisi\'on de Ciencias e Ingenier\'ias, Universidad de Guanajuato,\protect\\ Loma del Bosque 103, Lomas del Campestre CP 37150, Le\'on, Gto., M\'exico}
 \affiliation{Theoretical Physics Group and Quantum Alberta,
Department of Physics and Astronomy,
University of Lethbridge,
4401 University Drive, Lethbridge,
Alberta, T1K 3M4, Canada}
\author{Saurya Das}
 \email{saurya.das@uleth.ca}
\affiliation{Theoretical Physics Group and Quantum Alberta,
Department of Physics and Astronomy,
University of Lethbridge,
4401 University Drive, Lethbridge,
Alberta, T1K 3M4, Canada}

\date{\today}

\begin{abstract}
The Generalized Uncertainty Principle and the related minimum length are normally considered in non-relativistic Quantum Mechanics.
Extending it to relativistic theories is
important for having a Lorentz invariant minimum length and for testing the modified
Heisenberg principle at high energies.
In this paper, we formulate a relativistic
Generalized Uncertainty Principle. We then use this
to write the modified
Klein-Gordon,
Schr\"odinger
and Dirac equations,
and compute quantum gravity corrections to
the relativistic hydrogen atom, particle in a box, and the linear harmonic oscillator.
\end{abstract}
\pacs{Valid PACS appear here}
\maketitle









\section{\label{sec:Int}Introduction}
Theories of Quantum Gravity, such as String Theory \cite{Physics1988,AHARONY2000183,Magueijo:2004vv}, Loop Quantum Gravity \cite{Rovelli:2010vv,Rovelli:2010wq,Smolin:2004sx,Antonio2018} as well as Doubly-Special Relativity theories \cite{Amelino-Camelia2010,Smolin2003}, have one thing in common:
they all predict a minimum measurable length or a scale in spacetime.
This is in direct contradiction with the Heisenberg uncertainty principle \cite{Heisenberg1927}, since the latter allows for infinitely small uncertainties in position.
Phenomenological approaches to Quantum Gravity can describe the existence of such a minimal length via the so-called Generalized Uncertainty Principle (GUP)
\cite{Kempf1995,Ali2011,Bosso2018,Ali2010,Das2011,Das2008,Das2014}.
While the minimum length, normally considered to be the
Planck length, $\ell_{Pl}=10^{-35}$ m, signifies the scale at which quantum gravity effects would become manifest, one is faced with the problem that theories incorporating a minimum length break Lorentz covariance, simply because a measured length is not a Lorentz invariant quantity.
This effectively chooses a special frame of reference,
bringing back the notion of an aether.
Because of this difficulty, GUP models so far have been mainly considered in non-relativistic theories
with few attempts in relativistic theories and quantum field theory
\cite{Hossenfelder2006,Kober2010,Pramanik2013_1,Pramanik2014_1,Capozziello:1999wx,Kober:2010sj,Kober:2011dn,Shibusa:2007ju,Husain:2012im,Zakrzewski1994,Hill2002}.
In the present work, we study this aspect and propose a Lorentz
covariant GUP which reduces to its familiar  non-relativistic versions at low energies, and incorporates a minimum length.

This paper is organized as follows.
In Section \ref{sec:Lorentz}, we start with the most general quadratic GUP (i.e. only quadratic terms in momenta), and do a relativistic generalization thereof, similar to \cite{Quesne2006}.
We proceed to calculate the Poincar\'e algebra associated with the newly defined position and momentum, along with the corresponding Casimir operators.
We then show that the corresponding algebras are unmodified for a certain range of parameters.
In Section \ref{sec:Applications}, using the modified dispersion relation, we compute GUP corrections to the
modified Klein-Gordon, relativistically corrected Schr\"odinger and the Dirac equations.
Finally, we apply them to a set of experiments to impose bounds on the GUP parameters.
In Section \ref{sec:Concl}, we present a summary of the results and a list future works.

\section{\label{sec:Lorentz}Relativistic Generalized Uncertainty Principle}

We start from the algebra for position $x^{i}$ and momentum $p^{i}$ proposed in \cite{Kempf1995},
giving us the well-known quadratic GUP
\begin{equation}
[x^{i},p^{j}]=i\hbar\delta^{ij}\,\left(1+f(\vec{p}\,^2)\right)\,.
\label{kmm1}
\end{equation}
In the above,
$i\,,j\in\{1,2,3\}$, and Eq.(\ref{kmm1}) is non-relativistic.
In \cite{Kempf1995}, the authors show that the position operators obey the following commutation relation
\begin{equation}
[x^i,x^j]=-i\hbar f^{\prime}(\vec{p}^2)\left(x^ip^j-x^jp^i\right)\,.
\end{equation}
Then, considering $f(\vec{p}^2)=\beta_1\vec{p}^2$, they proceed calculating the minimal uncertainty in position corresponding to that  uncertainty relation, finding $\Delta x_{\textit{min}}=\hbar\sqrt{\beta_1}$.
Inspired by \cite{Quesne2006}, we will expand this algebra to the full Minkowski spacetime, with the following signature  $\{-,+,+,+\}$.
In particular, we will consider the following commutator
\begin{equation}
\label{GUPxp}[x^{\mu},p^{\nu}]=i\hbar\,\left(1+(\varepsilon-\alpha)\gamma^2 p^{\rho}p_{\rho}\right)\eta^{\mu\nu}+i\,\hbar(\beta+2\epsilon)\gamma^2p^{\mu}p^{\nu}\,,\\
\end{equation}
where $\mu\,,\nu\in\{0,1,2,3\}$ and $\alpha$, $\beta$, $\varepsilon$, and $\xi$
are dimensionless parameters which can be used to fix the particular model
\footnote{The parameters used in \cite{Kempf1995} and \cite{Quesne2006} are related to the ones used here as follows:
 $\beta_1=(\alpha+\varepsilon)\gamma^2$ and $\beta_2=(\beta+2\varepsilon)\gamma^2$. }
.
We will assume the parameter $\gamma$, with dimensions of inverse momentum, to be $\gamma=\frac{1}{c\,M_{Pl}}$, where $M_{Pl}$ is the Plank mass.
In this way, the minimal length will be associated with the Planck length.
Note that Eq.\eqref{GUPxp} reduces in the non-relativistic ($c\rightarrow \infty$) limit to
Eq.(\ref{kmm1}) and in the non-GUP limit
($\gamma \rightarrow 0$) limit to the standard Heisenberg  algebra.
Note also that, while $x^{\mu}$ and $p^{\nu}$ are the physical position and momentum, they are not canonically conjugate.
Therefore, we introduce
two new $4$-vectors,
$x_0^\mu$ and $p_0^\nu$ which are
canonically conjugate, such that
\begin{align}
\label{MathVar}p_0^{\mu} = & -i\hbar\frac{\partial}{\partial x_{0\,\mu}}, &
[x_0^{\mu},p_0^{\nu}] = & i\hbar\eta^{\mu\nu}.
\end{align}\\
and in terms of which the physical position and momentum can be written, up to second order in $\gamma$, as follows
%
\begin{align}
\label{GUPx}x^{\mu}&=x_0^{\mu}-\alpha\gamma^2 p_0^{\rho}p_{0\rho}x_0^{\mu}+\beta\gamma^2p_0^{\mu}p_0^{\rho}x_{0\rho}+\xi\gamma^2 p_0^{\mu}, \\
\label{GUPp}p^{\mu}&=p_0^\mu\,(1+\varepsilon\gamma^2 p_0^{\rho}p_{0\rho})\,.
\end{align}
Using Eqs.(\ref{GUPx}) and (\ref{GUPp}),
we get
\begin{equation}
    \label{GUPxx}[x^{\mu},x^{\nu}]=
i\hbar\gamma^2\frac{2\alpha+\beta}{1+(\varepsilon-\alpha)\gamma^2 p^{\rho}p_{\rho}}\left(x^{\mu}p^{\nu}-x^{\nu}p^{\mu}\right).
\end{equation}
Therefore as in \cite{Kempf1995},
in this case too one arrives at a non-commutative spacetime.
We also notice that the above algebra does not close and as a result does not have a corresponding
Lie manifold.

Note also that the last two terms in Eq.\eqref{GUPx} break isotropy of spacetime
by introducing the preferred direction of $p_0^\mu$.
Since this violates the principles of relativity,
we will assume that $\beta=\xi=0$ from now on.

\subsection{\label{sec:LnP} Lorentz and Poincar\'e algebra}

 Using Eqs.\eqref{GUPx} and \eqref{GUPp}, we now construct the generators of the Lorentz group
\begin{equation}
    M^{\mu\nu} = p^{\mu}x^{\nu}-p^{\nu}x^{\mu}
    = \left[1+(\varepsilon-\alpha)\gamma^2p_0^{\rho}p_{0\,\rho}\right]\tilde{M}^{\mu\nu}\,,
\end{equation}
%
where $\tilde{M}^{\mu\nu}=p_0^{\mu}x_0^{\nu}-p_0^{\nu}x_0^{\mu}$ is the Lorentz generators constructed from the canonical variables  $x_0$  and $p_0$. Next we compute the Poincar\'e algebra for the physical operators
\begin{align}
  \nonumber  [x^\mu,M^{\nu\rho}] &=  i\hbar[1 + (\varepsilon - \alpha) \gamma^2 p^{\rho} p_{\,\rho}]\left(x^{\nu}\delta^{\mu\rho}-x^{\rho}\delta^{\mu\nu}\right)\\
 &\label{xM} + i\hbar 2 (\varepsilon - \alpha) \gamma^2 p^{\mu} M^{\nu\rho}\\
   \label{pM}[p^\mu, M^{\nu\rho}]& =  i\hbar[1 + (\varepsilon - \alpha) \gamma^2 p^{\rho} p_{\,\rho}]\left(p^{\nu}\delta^{\mu\rho}-p^{\rho}\delta^{\mu\nu}\right),\\
    \label{MM}\nonumber [M^{\mu\nu},M^{\rho\sigma}] &= i\hbar\left(1 + (\varepsilon - \alpha) \gamma^2 p^{\rho} p_{\,\rho}\right)\left(\eta^{\mu\rho}M^{\nu\sigma}\right.\\
    &-\eta^{\mu\sigma} M^{\nu\rho}-\left.\eta^{\nu\rho}M^{\mu\sigma}+\eta^{\nu\sigma}M^{\mu\rho}\right)\,.
\end{align}
Inspecting
Eqs.(\ref{xM},\ref{pM},\ref{MM}),
we see that on the line
$\varepsilon=\alpha$ in the parameter space, one has a non-trivial relativistic GUP with an {\it unmodified} Poincar\'e algebra.
We will restrict ourselves to
this choice for the rest of the article, for which the GUP algebra
and the non-commutativity of spacetime follows from
Eqs.(\ref{GUPxp}) and (\ref{GUPxx}), respectively:
\begin{align}
[x^{\mu},p^{\nu}]&=i\hbar\,\left(\eta^{\mu\nu}+2\alpha\gamma^2p^{\mu}p^{\nu}\right)\,, \label{up1} \\
[x^{\mu},x^{\nu}]&=-
2i\hbar\alpha\gamma^2\left(x^{\mu}p^{\nu}-x^{\nu}p^{\mu}\right)\,, \label{nc1}
\end{align}
where $\alpha>0$.
Notice that for any one-dimensional spatial component, Eq.(\ref{up1}) above reduces to Eq.(\ref{kmm1}) \cite{Kempf1995} and the quadratic part of Eq.(1) of \cite{Ali2011} (up to an unimportant numerical factor).
It is interesting to note that algebras
Eqs.(\ref{up1}) and (\ref{nc1})
have similarities to the ones proposed
in \cite{Snyder:1946qz}.
We will study further implications of the above in the following sections.

\section{\label{sec:Applications}Applications}

For the case
$\varepsilon=\alpha$, Eqs. (\ref{GUPx}) and (\ref{GUPp}) take the following form,
\begin{align}
\label{X}    x^{\mu}&=x_0^{\mu}(1-\alpha\gamma^2 p_0^{\rho}p_{0\rho})\\
    \label{P} p^{\mu}&=p_0^\mu\,(1+\alpha\gamma^2 p_0^{\rho}p_{0\rho})\,.
\end{align}
As we showed earlier, since now the the definition of $M^{\mu\nu}$ in terms of the physical position and momentum, $x^{\mu}$ and $p^{\mu}$, as well as the Poincar\'e algebra, remains unchanged, the squared physical momentum $p^{\rho}p_{\rho}$ is again a Casimir invariant, commuting with every other operator in the group.
Using this, we can derive the dispersion relation and the Klein-Gordon (KG) equation.
It takes the following form
\begin{equation}
    p^{\rho}p_{\rho}=-(mc)^2,
\end{equation}
or, in terms of the variables $p_0^{\mu}$,
\begin{equation}
    \label{ModKG}
    p_0^{\rho}p_{0\rho}(1+2\alpha\gamma^2p_0^{\sigma}p_{0\sigma})=-(mc)^2 \,,
\end{equation}
where $m$ is the mass of the particle.
Observe that using Eq.(\ref{MathVar}),
the KG equation now is a fourth order equation, with four
linearly independent solutions.
{Solving Eq.(\ref{ModKG}) for $p_0^{\rho}p_{0\rho}$,} we obtain
\begin{eqnarray}
\label{ModPar}
p_0^{\rho}p_{0\rho}&&=-\frac{1}{4\alpha\gamma^2}+\sqrt{\frac{1}{\left(4\alpha\gamma^2\right)^2}-\frac{(mc)^2}{2\alpha\gamma^2}}\ \\
&&
\simeq - ( m\,c)^2-2\alpha\gamma^2(m\,c)^4-{\cal O}\left(\gamma ^4\right)\,,\label{minus}
\end{eqnarray}
where we have discarded the other solution since it does not reduce to $(mc)^2$ in the $\gamma\rightarrow 0$ limit.
In this process, we also lose two remaining solutions of the fourth order equation (\ref{ModKG}).
As was shown in  \cite{Ali2011},
including those solutions introduced very small corrections and we will ignore them in this paper.
In the next three subsections, we study applications of the above equation as well as the GUP-modified Dirac equation for a number of quantum systems.

\subsection{Klein-Gordon equation}

Writing Eq.(\ref{minus}) as an operator equation on the wavefunction $\Psi$ and
using Eq.\eqref{MathVar}, we
obtain the following GUP-modifed Klein-Gordon equation:
%
%
%
%
%
%
\begin{equation}\label{ModKGDiff}
    \frac{1}{c^2}\frac{\partial^2}{\partial t_0^2}\Psi-\nabla_0^2\Psi+\frac{1}{\hbar^2}\left[  (m\,c)^2+2 \alpha  c^4 \gamma ^2 m^4 \right]\Psi=0\,.
\end{equation}
It is worth noticing that in the limit $\gamma\rightarrow 0$, the above equation reduces to the standard KG equation.
Moreover, the solutions of the modified equation have the same form of the standard one but with modified parameters.

\subsubsection{Energy spectrum for Relativistic Hydrogen atom}

Introducing the minimal coupling
\cite{Bosso:2018uus,Das:2009hs}
\begin{equation}\label{minimalcoupling}
    \frac{\partial}{\partial x_{0\rho}}\rightarrow \frac{\partial}{\partial x_{0\rho}}+ \frac{ie}{\hbar}\,A^{\rho}\,
    \end{equation}
and the Hydrogen atom nuclear potential in Coulomb gauge
%
%
%
%
%
\begin{equation}\label{coulomb}
    A_{\rho} = \left\{\frac{e}{4\pi\varepsilon_0\,r },0,0,0\right\},
\end{equation}
in Eq.(\ref{ModKGDiff}),
we obtain the GUP modified Klein-Gordon equation for the Hydrogen atom
%
\begin{align}\label{HA}
\nonumber &-\left( i\frac{\partial }{\partial t_0}+ c\frac{\kappa}{r}\right)^2\Psi-c^2\nabla_0^2\Psi \\
    &+\frac{c^2}{\hbar^2}\left[ (m\,c)^2+2\alpha\gamma^2(m\,c)^4 \right]\Psi=0,
\end{align}
In the above, $\kappa={e^2/4\pi\epsilon_0 \hbar c}$ is the fine structure constant.
The solution of Eq.\eqref{HA} in spherical coordinates is of the form
\cite{Ducharme2010}
\begin{equation}
    \Psi_{nlm}=R_{nl}(r)\,Y_{lm}(\theta,\phi)\,e^{-E_{0\,nl}t/\hbar}\,,
\end{equation}
%
%
where the $R_{nl}(r)$ are  spherical Bessel functions and $Y_{lm}(\theta,\phi)$ the spherical harmonics. The energy
levels for these solutions are given as
\begin{equation}\label{HAenergy}
    E_{0\,nl}=\left( m\,c^2+\alpha\gamma^2m^3c^4 \right)\left[1+\frac{\kappa^2}{(n+1-\eta)^2}\right]^{-1/2}\,,
\end{equation}
where
\begin{equation}\label{40}
    \eta=\frac{1}{2}\pm\sqrt{\left(l+\frac{1}{2}\right)^2-\kappa^2},
\end{equation}
where the minus sign recovers the standard
Hydrogen atom spectrum
from Eq.\eqref{HAenergy} in the non-relativistic limit, with the identification
$N\equiv n+1-\eta$ for the the hydrogen atom energy level.
Expanding Eq.\eqref{HAenergy} in powers of $\kappa^2$, we find
\begin{align}\label{E_0KG}
\nonumber E_{0\,N}=&\left(m\,c^2+ \alpha  \gamma ^2  m^3c^4 \right)-\frac{ \kappa ^2 \left(m\,c^2 \right)}{2N^2}\\&
+\frac{3 \kappa ^4  \left(mc^2\right)}{8N^4}+\frac{ 3\kappa ^4  \left(\alpha\gamma^2m^3c^4\right)}{8N^4}-\frac{ \kappa ^2 \left(\alpha\gamma^2m^3c^4\right)}{2N^2}.
\end{align}

We now translate these corrections to $E_0$ (the $0^{th}$ component of $p_0^\mu$)
to corrections to
$E$ (the $0^{th}$ component of $p^\mu$ or the physical momentum),
using the appropriate component of %
Eq.\eqref{P}, namely
\begin{align}
  \nonumber  E =& E_0\left(1+\alpha\gamma^2 p_0^\rho p_{0\,\rho}\right)\\
    =& E_0 \left[1 - \alpha \gamma^2 (mc)^2 \right] + {\cal O}(\gamma^4),\label{PhysEnrg}
\end{align}
to obtain
\begin{align}\label{EnergyWithCorrections}
 E_N=&\left(m\,c^2
\right)-\frac{ \kappa ^2 \left(m\,c^2 \right)}{2N^2}
+\frac{ \kappa ^4  \left(mc^2\right)}{8N^4}
\end{align}
As evident from the above equation, all the  GUP corrections of $\gamma^2$ vanish.
It is important to clarify that the Casimir still remains modified which will give modified results in field theory.
Furthermore, including the remaining two solutions of the fourth order equation (\ref{ModKG}) may give rise to GUP corrections.
We will investigate this elsewhere

\subsection{Schr\"odinger equation with relativistic and GUP corrections}
We can write Eq. \eqref{minus} as
\begin{equation}
    -E_0^2+c^2\vec{p}_0^2+(mc^2)^2+2\alpha\gamma^2m^4c^6=0.
\end{equation}
We now expand it to fourth order in $\vec{p}_0$ and second order in $\gamma$ to obtain
\begin{align}\label{ModSEp}
\nonumber E_0=&mc^2\left(1+\alpha\gamma^2(mc)^2\right)+\frac{\vec{p}_0^2}{2m}\left(1-\frac{1}{2}\alpha\gamma^2(mc)^2\right)\\&-\frac{\vec{p}_0^4}{8m^3c^2}\left(1-3\alpha\gamma^2(mc)^2\right)\,.
\end{align}
Next, using Eq. \eqref{PhysEnrg}, we get the expression for the physical energy
\begin{multline}
    E = mc^2 + \frac{\vec{p}_0^2}{2m} \left[1 - \frac{3}{2} \alpha \gamma^2 (mc)^2\right] \\
    - \frac{\vec{p}_0^4}{8 m^3 c^2}\left[1 - 4 \alpha \gamma^2 (mc)^2\right]\,,
\end{multline}
which consists of the rest mass,
non-relativistic kinetic energy, relativistic and GUP corrections.

Next, the operator version of Eq.(\ref{ModSEp}), using $E_0 = i \hbar \frac{\partial}{\partial t_0}$ and $\vec{p}_0 = - i \hbar \vec\nabla_0$, and including a potential $V (\vec x)$, yields the modified Schr\"odinger equation with relativistic and GUP corrections
\begin{multline}
    i \hbar \frac{\partial}{\partial t_0} \Psi(t_0,\vec{x}_0)
    = \left[mc^2 \left(1 + \alpha \gamma^2 m^2 c^2 \right) \right. \\
    + \frac{(- i \hbar)^2}{2 m} \left(1 - \frac{1}{2}\alpha \gamma^2 m^2 c^2 \right) \nabla_0^2 \\
    \left. - \frac{(-i\hbar)^4}{8 m^3 c^2}\left(1 - 3 \alpha \gamma^2 m^2 c^2 \right) \nabla_0^4 + V(\vec{x})\right] \Psi(t_0,\vec{x}_0)\,.
\end{multline}
Omitting the rest energy term, we apply the above equation to a couple of
problems.

\subsubsection{Corrections for particle in a box}
Using Eq.\eqref{ModSEp}, we can write the Schr\"odinger equation with relativistic and GUP corrections.
Let us first consider a $1+1$-dimensional case with the following potential
\begin{equation}
    V(x)=\left\{
    \begin{array}{lll}
        V_0 & \text{for} & 0<x<L,\\
        \infty & \text{for} & x \leq 0 ~ \cup ~ x\geq L.\\
    \end{array} \right.
\end{equation}
Including the potential in Eq.\eqref{ModSEp}, we
obtain the Schr\"odinger equation for one dimensional particle in a box plus small perturbations.

Consider the wave function for the unperturbed ($\gamma = 0,c\rightarrow \infty$) case
\begin{equation}
    \Psi_n(t_0,\vec{x}_0)=\sqrt{\frac{2}{L_0}}\sin\left(\frac{n\pi}{L_0}x_0\right)e^{-i\omega_n\,t_0}\,.
\end{equation}
From Eq.\eqref{X},
one can read-off the physical dimensions of the box to be
\begin{equation}
    L=L_0(1+\alpha\gamma^2 (m\,c)^2+{\cal O}(\gamma^4) )\,.
\end{equation}
The corrected  spectrum of $E_0$ is
%
%
%
\begin{multline}\label{Pbox}
    E_{0n} =  
    - \frac{1}{2m} \left(\frac{n \pi \hbar}{L_0}\right)^2 \left[1 - \frac{1}{2} \alpha \gamma^2 (mc)^2\right]\\
    + \frac{\hbar^4}{8 m^3 c^2} \left(\frac{n \pi}{L_0}\right)^4 \left[1 - 3 \alpha \gamma^2 (mc)^2 \right]\,,
\end{multline}
%
%
Using Eq.\eqref{PhysEnrg}, we translate this to the following expression for the physical energy
\begin{multline}
    E_{n} =
   - \frac{1}{2m} \left(\frac{n \pi \hbar}{L}\right)^2 \left[1 + \frac{3}{2} \alpha \gamma^2 (mc)^2\right]\\
   - \frac{\hbar^4}{8 m^3 c^2} \left(\frac{n\pi}{L}\right)^4.
\end{multline}
The first term corresponds to the non-relativistic energy with GUP-corrections, while the last to relativistic corrections.

The results of this section can be applied to experiments measuring directly the energy levels of quantum dots.
Comparing GUP corrections to the unperturbed energy term found above and equating to the accuracy of experiments measuring the energy levels of single quantum dot 
\cite{Hill2002}, we get
\begin{equation}
    \frac{\Delta E_{n}}{E_{n}} = \frac{2}{3}\alpha\gamma^2m^2c^2 \sim \alpha 10^{-42} \lesssim 10^{-1}~.
\end{equation}
From this, one gets an upper bound on $\alpha$
\begin{equation}
    \alpha \lesssim 10^{41}\,.
\end{equation}

\subsubsection{Corrections to the linear harmonic oscillator }

Next, we include the harmonic oscillator potential in  Eq.\eqref{ModSEp}
\begin{equation}
V(x)=\frac{1}{2}m\,\omega^2\,x^2\,.
\end{equation}
This gives rise to the following $E_0$
%
\begin{multline}
    E_0 = \frac{p_0^2}{2m} + \frac{1}{2} m \omega^2 x_0^2 [1 + 2 \alpha \gamma^2 (m c)^2]
    \\
    - \frac{p_0^2}{4m} \left(\alpha \gamma^2 m^2 c^2\right) - \frac{p_0^4}{8 c^2 m^3} + \frac{3p_0^4}{8m} \alpha \gamma^2.
\end{multline}
%
%
%
%
%
For the first order correction in perturbation theory, we obtain
\begin{multline}
    E_{0n} =
    \hbar \omega \left(n + \frac{1}{2}\right) \left(1 + \frac{1}{2} \alpha \gamma^2 m^2 c^2 \right)\\
    - \frac{\hbar^2 \omega^2}{32 m c^2} \left(1 - 3 \alpha \gamma^2 m^2 c^2\right) \left[5 n (n+1) + 3\right].
\end{multline}
From this, using Eq.\eqref{PhysEnrg}, we find for the physical energy
\begin{multline}
    E_{n} =
    \hbar \omega \left(n + \frac{1}{2}\right) \left(1 - \frac{1}{2} \alpha \gamma^2 m^2 c^2 \right)\\
    - \frac{\hbar^2 \omega^2}{32 m c^2} \left(1 - 4 \alpha \gamma^2 m^2 c^2\right) \left[5 n (n+1) + 3\right].
\end{multline}

Similar to the calculations for Landau levels done in \cite{Das2008},
we use experiments to put bounds on $\alpha$ 
%
%
%
\begin{equation}
    \frac{ \Delta E_{n}}{E_{n}}=-\frac{3}{4}\alpha\gamma^2m^2c^2\sim \alpha\,10^{-44}~.
\end{equation}
Equating this to the accuracy of direct measurements of Landau levels
\cite{PhysRevB.93.125422,PhysRevB.55.R16013}, we get
\begin{equation}
   \alpha\,10^{-44}\leq 10^{-3}\Rightarrow\alpha\leq 10^{41}.
\end{equation}
It is worth noting that this is several orders smaller than
that was obtained in
\cite{Das2008}.

\subsection{\label{Section:Dirac}Dirac equation and GUP corrections}

Starting from Eq.\eqref{ModPar}, working in the following signature $\{+,-,-,-\}$ signature, and considering the following Dirac matrices
\begin{align}
\tau^0=\begin{pmatrix}
\mathbf{I}&0 \\
0 & -\mathbf{I}
\end{pmatrix},\,\,\,\,\,\tau^i=\begin{pmatrix}
0&\sigma^i \\
\sigma^i &0
\end{pmatrix}\,,
\end{align}
%
%
we get the following GUP-modified Dirac equation
\begin{equation}
i\hbar\tau^{\mu}\frac{\partial}{\partial x_0^{\mu}}\Psi- \left(\tau^0\sqrt{\frac{1}{4\alpha\gamma^2}-\sqrt{\frac{1}{\left(4\alpha\gamma^2\right)^2}-\frac{(mc)^2}{2\alpha\gamma^2}}}\right)\Psi=0\,,
 \end{equation}
Truncating to ${\cal O}(\gamma^2)$, we get
\begin{equation}\label{ModD}
   i\hbar\tau^{\mu}\partial_{\mu}\Psi-\tau^0(mc+\alpha\gamma^2m^3c^3)\Psi=0.
\end{equation}
%

\subsubsection{Hydrogen atom}
%
%
%

Using Eqs.(\ref{minimalcoupling}) and (\ref{coulomb}) for minimal coupling and Coulomb potential, from Eq.\eqref{ModD} we get
\begin{equation}
    \left[i \tau^{\mu} \frac{\partial}{\partial x_0^{\mu}} - \frac{e}{\hbar} \tau^{\mu} A_{\mu} - \tau^0 \frac{(mc + \alpha \gamma^2 m^3 c^3)}{\hbar}\right]\Psi = 0,
\end{equation}
with the solution
\begin{equation}
    \Psi(t,r,\theta,\phi)=T(t)\frac{1}{r}\left(\begin{array}{cc}
    F(r)Y_{jm}  (\theta,\phi)     \\
           i\,G(r)Y'_{jm}  (\theta,\phi)
    \end{array}\right)\,.
\end{equation}
In the above, $F(r)$ and $G(r)$ are spherical Bessel functions, and $Y_{jm}$ and $Y'_{jm}$ are the spherical spinors.
Following \cite{Citation2017}, we find the energy spectrum
\begin{equation}
    E_{0nj}=(mc^2+\alpha\gamma^2m^3c^4)\left[1+\frac{\kappa^2}{(n-j-\frac{1}{2})^2}\right]^{-1/2}\,,
\end{equation}
where again $\kappa=e^2/4\pi\varepsilon_0\hbar c$ is the fine structure constant. Expanding the above equation in Taylor series in $\kappa$, we get
\begin{align}
  \nonumber E_{0N}= &(mc^2+\alpha\gamma^2m^3c^4)-\frac{ mc^2\kappa ^2}{2 N^2}-\frac{\alpha\gamma^2m^3c^4\kappa ^2}{2 N^2}\\
  &+\frac{3 mc^2 \kappa ^4}{8 N^4}+\frac{\alpha\gamma^2m^3c^4 \kappa ^4}{8 N^4}\,.
\end{align}
%
As before, using
Eq.(\ref{PhysEnrg}), we get
for the physical energy spectrum
$E_N$
\begin{align}
    E_{N}= mc^2
    -\frac{ mc^2\kappa ^2}{2 N^2}
  +\frac{3 mc^2 \kappa ^4}{8 N^4}\,,
\end{align}
Once again, we can see that the GUP modifications vanish for the physical energy.
%
%
%
%
%
%
%
%
\section{\label{sec:Concl}Conclusion}

In this article, we proposed a method of incorporating minimal length in relativistic quantum mechanics.
We achieved this by expanding proposing
a Generalized Uncertainty Principle to include both space and time.
As a result, we obtained a non-commutative spacetime.
%
For a set of GUP parametes, the Poincar\'e algebra remained unchanged, albeit with a modified Casimir operator.
We use this Casmir to
write down GUP-modfied quantum mechanical wave equations and applied them to
several examples estimate upper bounds on the GUP parameter.
We found GUP corrections in the case of the relativistically corrected Schr\"odinger equation, but did not find GUP corrections in the few applications of the Klein-Gordon and Dirac equations. We believe GUP corrections will result in the case of the last two for other problems. Furthermore,
this work is an important step in the application of a covariant GUP to Quantum Field Theory. The GUP-modified actions therein
will predict corrections to various scattering amplitudes .
We hope to report on this elsewhere.

\vspace{0.2cm}
\begin{center}
{\bf Acknowledgment}
\end{center}

We thank J. Stargen for discussions.
This work is supported by the Natural Sciences and Engineering Research Council of Canada, the University of Lethbridge and the Universidad de Guanajuato.

%




\end{document}